\documentclass[conference]{IEEEtran}

\IEEEoverridecommandlockouts
\usepackage{cite}
\usepackage{amsmath,amssymb,amsfonts}
\usepackage[T1]{fontenc}
\PassOptionsToPackage{bookmarks={false}}{hyperref}
\usepackage{algorithm}
\usepackage{algpseudocode}
\usepackage{graphicx}
\usepackage{tikz}

\usepackage{booktabs}
\usepackage{textcomp}
\usepackage{physics}
\usepackage{xcolor}
\usepackage[footnotesize]{caption}
\usepackage{subcaption}
\DeclareCaptionLabelFormat{bold}{\textbf{(#2)}}
\usepackage{multirow}
\usepackage{tikz}
\usepackage{orcidlink}
\usepackage{comment}
\usepackage{amsmath}
\usepackage{amsfonts}
\usepackage{amssymb}
\def\BibTeX{{\rm B\kern-.05em{\sc i\kern-.025em b}\kern-.08em
    T\kern-.1667em\lower.7ex\hbox{E}\kern-.125emX}}    
\begin{document}
\begin{NoHyper}
\title{FedQNN: Federated Learning using Quantum Neural Networks}

\author{\IEEEauthorblockN{Nouhaila Innan\textsuperscript{1}, Muhammad Al-Zafar Khan\textsuperscript{2}, Alberto Marchisio\textsuperscript{3,4}, Muhammad Shafique\textsuperscript{3,4}, and\\
Mohamed Bennai\textsuperscript{1}}
\IEEEauthorblockA{\textsuperscript{1}Quantum Physics and Magnetism Team, LPMC, Faculty of Sciences Ben M'sick, Hassan II University of Casablanca,\\
Morocco\\
\textsuperscript{2}Quantum United Arab Emirates, UAE\\
\textsuperscript{3}eBRAIN Lab, Division of Engineering, New York University Abu Dhabi (NYUAD), Abu Dhabi, UAE\\
\textsuperscript{4}Center for Quantum and Topological Systems (CQTS), NYUAD Research Institute, NYUAD, Abu Dhabi, UAE\\
nouhaila.innan-etu@etu.univh2c.ma, m.khan@quae.ae, alberto.marchisio@nyu.edu, muhammad.shafique@nyu.edu,\\
mohamed.bennai@univh2c.ma
}}

\maketitle

\begin{abstract}
In this study, we explore the innovative domain of Quantum Federated Learning (QFL) as a framework for training Quantum Machine Learning (QML) models via distributed networks. Conventional machine learning models frequently grapple with issues about data privacy and the exposure of sensitive information. Our proposed Federated Quantum Neural Network (FedQNN) framework emerges as a cutting-edge solution, integrating the singular characteristics of QML with the principles of classical federated learning. This work thoroughly investigates QFL, underscoring its capability to secure data handling in a distributed environment and facilitate cooperative learning without direct data sharing. Our research corroborates the concept through experiments across varied datasets, including genomics and healthcare, thereby validating the versatility and efficacy of our FedQNN framework. The results consistently exceed 86\% accuracy across three distinct datasets, proving its suitability for conducting various QML tasks. Our research not only identifies the limitations of classical paradigms but also presents a novel framework to propel the field of QML into a new era of secure and collaborative innovation.
\end{abstract}


\section{\label{sec:level1}Introduction}

Noisy Intermediate-Scale Quantum (NISQ) devices have brought about new possibilities and challenges in Quantum Machine Learning (QML) \cite{refa,refb}; despite their potential, these devices need to be improved to mitigate inherent quantum noise, which limits the large-scale applicability of QML algorithms. 
One of the key hurdles in implementing QML is the computational expense of training Quantum Neural Networks (QNNs) and their associated quantum layers \cite{refc, refd, refd1, refd2,refd3, refd4, refd5}. This challenge mirrors the difficulties in training Deep Neural Networks (DNNs). The concept of Federated Machine Learning (FedML) has been introduced to address this in the classical realm of ML, which represents a paradigm shift in model training and deployment, addressing key concerns in big data and privacy-sensitive applications. The concept of FedML was pioneered by MacMahan et al. at Google, who introduced a decentralized model building procedure to overcome the drawbacks associated with centralized data processing, such as privacy risks and hardware limitations \cite{ref5}.

FedML operates on distributive training, where models are trained on localized datasets across multiple nodes or ``clients'' without sharing the actual data between them, thus ensuring privacy preservation. The model parameters are then aggregated to create a comprehensive model for end-users. This decentralized training approach enhances data privacy and reduces the computational and hardware resources required for model training. The advantages of FedML include:
\begin{itemize}
\item Non-exposure of sensitive data between clusters.
\item Reduced computational and hardware resources due to distributed model training.
\item Access to a heterogeneous dataset from various industry sectors increases the scope and diversity of the models.
\end{itemize}
Building upon the principles of FedML, Quantum Federated Learning (QFL) emerges as a specialized solution for the quantum domain \cite{refe,reff}. QFL adapts the federated learning framework to the unique characteristics and requirements of Quantum Computing (QC). This adaptation is essential to address challenges such as the fragility of quantum states and to integrate noise mitigation strategies, which are crucial in the NISQ era.

However, the integration of FedML into QC is more complex. The distinctions between FedML and QFL are significant, necessitating thorough understanding and adaptation, which includes handling quantum data and establishing communication protocols between quantum and classical nodes. Quantum data, represented by superimposed qubit states, are inherently fragile due to decoherence and susceptibility to noise. This fragility requires sophisticated quantum error correction and noise mitigation strategies, distinct from classical FedML approaches. Conversely, classical FedML has advanced significantly, with many contemporary-developed optimization procedures that have enhanced its efficiency and robustness. This maturity contrasts sharply with the nascent stage of QFL, which is still primarily grappling with the complexities of quantum data and the nuanced challenges of integrating quantum and classical systems. Therefore, while classical FedML strides ahead with well-established protocols and optimizations, QFL is navigating its foundational phase, focusing on addressing the fundamental challenges unique to QC and learning.

Additionally, the communication between quantum and classical nodes requires novel protocols to translate Quantum Information (QI) into a classical form and vice versa. This is a challenging task due to the quantum-classical information barrier and asymmetry. Existing QFL methods face several challenges, such as the efficient and secure transmission of QI \cite{refc1}, stabilization of quantum states during learning \cite{refy}, and scalability of quantum federated networks \cite{ref2}. Despite these challenges, the motivation for pursuing QFL is compelling, offering the potential to leverage QC's power while maintaining the privacy and decentralized nature of FedML. This combination could lead to significant advancements in finance \cite{ref9, ref10}, condensed matter physics and chemistry \cite{ref11}, healthcare \cite{reftt,reftt2}, bioinformatics \cite{ref12}, tomography \cite{ref14} and more, where large-scale, privacy-preserving computational power is paramount.

\textbf{The novel contributions of this work can be summarized as follows:}
\begin{itemize}
\item Developing a QNN model that enables the creation of a quantum model precisely tailored to specific requirements, such as high-dimensional data processing, while ensuring robustness against quantum noise and adaptability to different quantum hardware architectures.
\item Constructing a novel \textit{Federated Quantum Neural Network (FedQNN)} framework with decentralized quantum model training and secure model update aggregation, leveraging the strengths of collaborative learning while ensuring data privacy.
\item Conducting comprehensive experiments on three diverse datasets: Iris, breast cancer, and a custom synthetic DNA dataset. These experiments demonstrate the broad applicability and versatility of the approach in genomics and medicine.
\item Exploring the impact of client numbers on the performance of our FedQNN method, providing valuable insight into the scalability and accuracy of the approach.
\item Evaluating our FedQNN framework on real Quantum Processing Units (QPUs) from IBM Quantum. This evaluation, using the synthetic DNA dataset, achieves an accuracy of more than 80\%, underscoring the practical effectiveness of the approach.
\end{itemize}
Through these research contributions, our FedQNN framework exhibits adaptability to diverse datasets, including those in healthcare and genomics, and demonstrates robust performance across different QPUs. \textit{Our objective is to accelerate the development of scalable and robust QML through QFL}, paving the way for wider adoption and real-world applications of this revolutionary technology. In addressing the specific challenges of QFL, \textit{our framework ensures efficient and secure transmission of QI by keeping local datasets on client devices and employing secure communication for model updates}. This strategy helps in stabilizing quantum states during learning, as sensitive states are not transferred over the network, but rather, only model updates are aggregated. Furthermore, the decentralization of the learning process significantly improves the scalability of our framework, allowing for more clients to be accommodated without a loss in performance or security. 

This paper is organized as follows. In Sec.~\ref{sec2}, we provide a comprehensive background on FedML and a literature review of the most relevant works in the field. In Sec.~\ref{sec3}, we discuss our FedQNN framework and the algorithmic approach we have employed. In Sec.~\ref{sec4}, we present the results of our experiments. In Sec.~\ref{sec3}, we conclude this research and reflect on the results.

\section{Background and Related Work \label{sec2}} 

\subsection{QC and QML}
Classical physics laws break down on the subatomic level due to measurement uncertainty and the observation of weird particle behaviors. Pondering this thought, the Nobel laureate Richard Feynman \cite{ref100} conceived the idea of using a computer to simulate quantum systems in a way that holistically leveraged the principles of quantum mechanics for computing. As opposed to storing information in classical $0,1$ bits, information can be stored as superimposed states of $\ket{0}$ and $\ket{1}$ due to quantum coherence; \textit{qubits} -- The weaving together of ``quantum'' and ``bit''. Allowing states to exist in multiple states simultaneously opens up unforeseen capabilities for computing and information storage. However, QC technology is still in its infancy in terms of development, as the current state-of-the-art quantum computers are plagued by environmental noise and decoherence.     

QML is the coalescence and enhancement of regular ML with QC. A typical QML model that is ubiquitously used, as presented in Fig. \ref{generalqml}, follows the parameter-adjustment paradigm, works as follows: Classical features $\mathbf{x}=\left(x_{1},x_{2},\ldots, x_{m}\right)\in\mathbb{R}^{k\times m}$ are mapped into quantum states, $\left\{\ket{\psi_{i}}\right\}_{i=1}^{n}$ via some encoding scheme; $\varphi:\mathbf{x}\to\ket{\psi}$. Often, this involves data preprocessing and feature engineering, and since there are a limited number of qubits to work with \cite{qubit}, the number of quantum states is less than the number of classical data points, i.e., $\text{card}(\ket{\psi_{i}})\leq n$. The quantum states are then fed to a gate model that acts like a feed-forward neural network: $f(H^{\otimes n}, X, Y, R_{\chi}, \cdots)$, which creates superposition amongst the qubit states, and performs a series of operations to adjust the network parameters to attain optimality. Measurements on the states are then performed, and the criterion for checking whether an optimal solution is achieved is the loss function, $J(y,\hat{y};\boldsymbol{\omega})\approx 0$ and subsequently obtaining the weights $\boldsymbol{\omega}^{*}=\left(\omega_{1}^{*},\omega_{2}^{*},\ldots, \omega_{p}^{*}\right)=\arg\min_{\boldsymbol{\omega}}\; J$; if optimality is not achieved, the network then iteratively adjusts the parameters, and if it is, then a prediction, $\hat{y}$, is made. 
\begin{figure}
    \centering
    \includegraphics[width=1\linewidth]{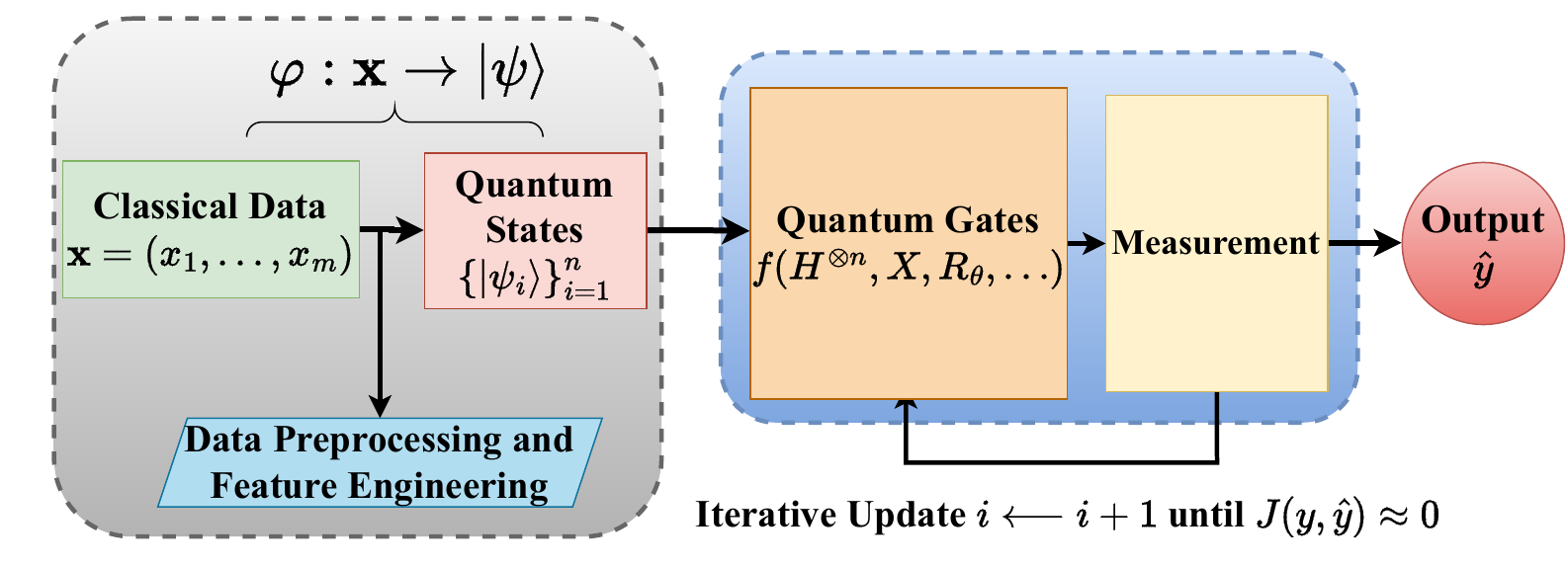}
    \caption{The general architecture of QML models starts with classical data $\mathbf{X}$, which undergoes data preprocessing and feature engineering. This data is converted into quantum-compatible states ${|\psi_j\rangle}$ via a mapping function $\phi$. For computation, these states are processed by a quantum circuit using quantum gates $f(\theta_n, X, R_{\ldots})$, including rotation and entangling gates. The model includes an iterative update loop where parameters are refined based on the lost function $J(\psi, y)$ until nearly converging ($J(\psi, y) \approx 0$). Finally, the quantum circuit's output is measured to produce the final model output $g$ for various predictive tasks.}
    \label{generalqml}
\end{figure}

QML offers several benefits over classical ML. These include, amongst others:
\begin{enumerate}
\item The potential to solve certain classical problems exponentially faster than the current state-of-the-art methods. This speedup has already been realized in problems ranging from unstructured search to optimization problems.
\item Faster and more streamlined processing of data than classical processing methods. This is due to the Hilbert space being larger and more encompassing than classical real Euclidean spaces; $\dim\mathcal{H}>\dim\mathbb{R}$.
\item Some QML algorithms have been theoretically demonstrated to solve certain classes of problems in fewer computational steps when compared to their classical counterparts.
\end{enumerate}
By utilizing the quantum mechanical properties of superposition and entanglement, QML offers a promising alternate approach to classical ML with the potential to enhance and augment predictive and prescriptive analytical models. 
\subsection{FedML and QFL}
FedML is an archetype of ML whereby the same models are trained on localized datasets (distributive training) without the dataset of each local cluster being exposed to the other clusters (privacy preservation) \cite{ref5}, and then the model parameters are aggregated to give a holistic model that end-users can utilize for predictive analytics in a particular industry, or use-case.
Algorithm \ref{fedml} and  Fig. \ref{figfedml} highlight the key steps and components of the FedML process.
\begin{algorithm}[htpb]
\caption{FedML}
\label{fedml}
\begin{algorithmic}[1]
\State \textbf{Input:} Set of clients $\{\mathcal{C}_{1}, \mathcal{C}_{2}, \ldots, \mathcal{C}_{k}\}$, their corresponding datasets $\{\mathcal{D}_{1}, \mathcal{D}_{2}, \ldots, \mathcal{D}_{k}\}$, learning rate $\alpha$
\State \textbf{Output:} Aggregated model $\overline{\mathcal{M}}$

\While{$J(\boldsymbol{\theta}_{i,j};\mathcal{D}_{i})\not\approx 0$}
    \For{each client $\mathcal{C}_{i}$ in 1 \textbf{to} $k$}
        \State Train models $\mathcal{M}_{i}$ on datasets $\mathcal{D}_{i}$, WLOG do not share $\left(\mathcal{M}_{i},\mathcal{D}_{i}\right)$ with $\left(\mathcal{M}_{j},\mathcal{D}_{j}\right)$
    \EndFor
    \State Expose model parameters/weights
    \begin{equation*}  \boldsymbol{\theta}^{i}=\left(\theta^{1}_{1},\theta^{2}_{2},\ldots,\theta^{k}_{p}\right)
    \end{equation*}
    to centralized server
    \State Create the aggregated model 
    \begin{equation*}
    \overline{\mathcal{M}}=\text{agg}(\mathcal{M}_{1},\mathcal{M}_{2},\ldots,\mathcal{M}_{k}) \end{equation*} with parameters $\boldsymbol{\Theta}=\left(\phi_{1},\phi_{2},\ldots,\phi_{p}\right)$
    \State Update parameters: $\boldsymbol{\theta}_{i,j}\longleftarrow\boldsymbol{\theta}_{i,j}-\alpha\boldsymbol{\nabla}_{\boldsymbol{\theta}_{i,j}}J(\boldsymbol{\theta}_{i,j};\mathcal{D}_{i})$
\EndWhile
\State Expose $\overline{\mathcal{M}}$ to the public
\end{algorithmic}
\end{algorithm}
\begin{figure}[htpb]
\centering
\includegraphics[width=1.1\linewidth]{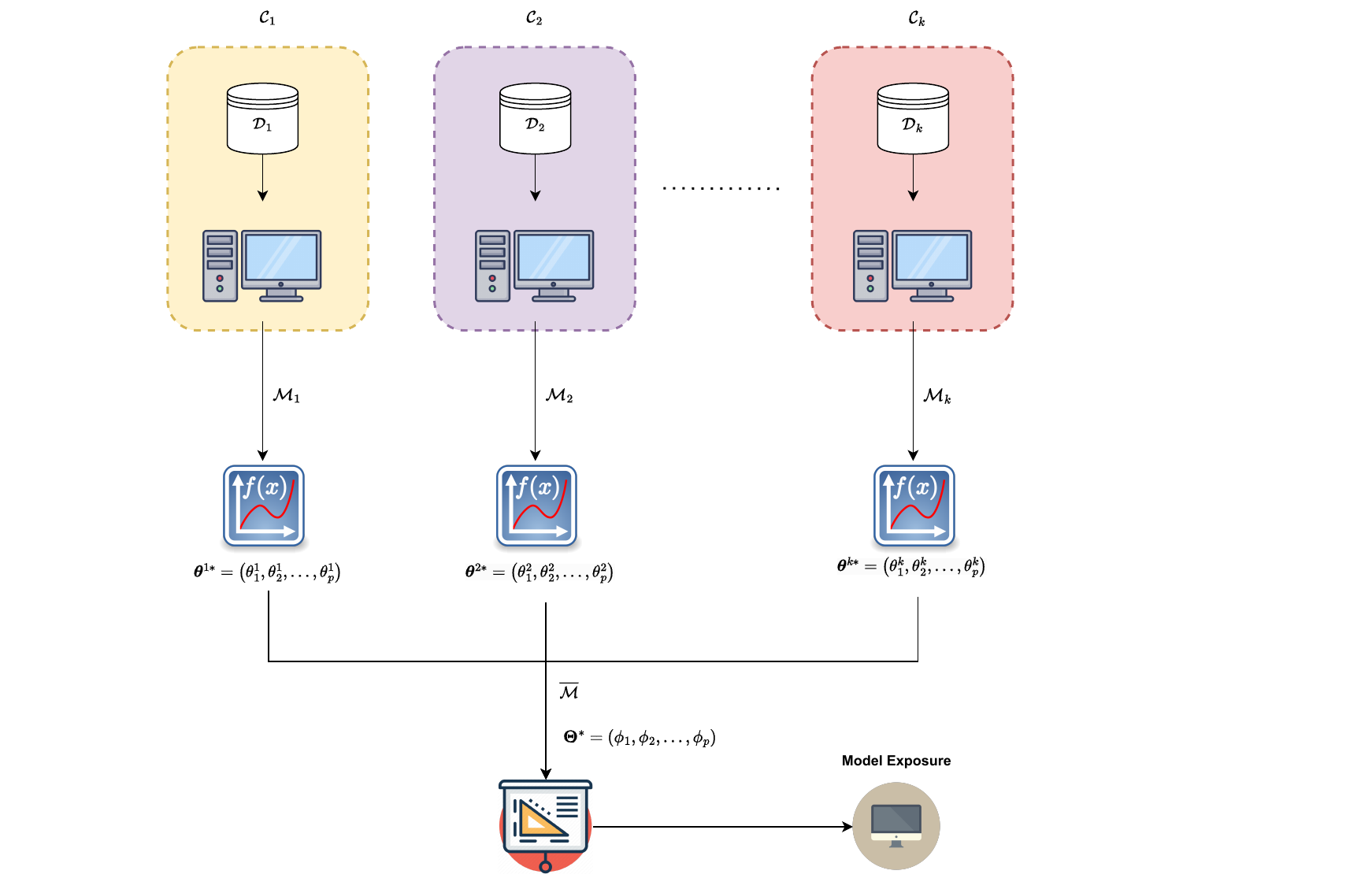}
    \caption{Network architecture for a typical FedML setup. Clients $\mathcal{C}_{i}$, for $1\leq i\leq k$, on local datasets $\mathcal{D}_{1},\mathcal{D}_{2},\ldots,\mathcal{D}_{k}$ with models $\mathcal{M}_{1},\mathcal{M}_{2},\ldots,\mathcal{M}_{k}$. The corresponding model parameters $\boldsymbol{\theta}^{1},\boldsymbol{\theta}^{2},\ldots,\boldsymbol{\theta}^{k}$, are sent to a centralized aggregation server, and an agglomerated model $\overline{\mathcal{M}}$ is made, and the model parameters $\boldsymbol{\Theta}^{}$ are exposed. The process is repeated until the model parameters are optimal.}
    \label{figfedml}
\end{figure}

QFL is analogous to FedML in computational steps, except it accounts for noise, and a very selective aggregation method needs to be employed not to introduce errors, and to minimize information loss. Common methods include:
\begin{enumerate}
\item \textbf{Riemannian Averaging:} Given $n$ quantum states $\ket{\psi_{1}},\ket{\psi_{2}},\ldots,\ket{\psi_{n}}$ with associated density matrices $\left\{\rho_{1},\rho_{2},\ldots,\rho_{n}\right\}\in\varrho$, the density operators are averaged according to
\begin{equation}
\bar{\rho}=\underset{\rho\in\varrho}{\arg\min}\sum_{i=1}^{n}d^{2}(\rho,\rho_{i}),
\end{equation}
where $d(\rho,\rho_{i})$ is the metric (distance) between $\rho$ and $\rho_{i}$.
\item \textbf{Schatten Norm Averaging:} Given $n$ quantum states $\ket{\psi_{1}},\ket{\psi_{2}},\ldots,\ket{\psi_{n}}$ with associated density matrices $\left\{\rho_{1},\rho_{2},\ldots,\rho_{n}\right\}\in\varrho$, the density operators are averaged according to
\begin{equation}
\bar{\rho}=\underset{\rho\in\varrho}{\arg\min}\sum_{i=1}^{n}||\rho-\rho_{i}||_{q},
\end{equation}
where 
\begin{equation}
||\rho||_{q}=\left(\sum_{i=1}^{n}\lambda_{i}^{q}\right)^{1/q},    
\end{equation}
$\lambda_{i}$ is the $i^{\text{th}}$ singular value of of $\rho$, and $||\rho||_{q}$ denotes the $q$-norm of the density operator. 
\end{enumerate}
Several other methods may be used to average the density operators; these include Krylov subspace averaging, kernel mean embeddings, ensembling, and many others. It would be an exercise in futility to repeat Algorithm \ref{fedml} for QFL. However, one should note that density matrices are associated with each quantum state, and the classical averaging methods should be avoided. 
Below, we review related work.

In \cite{ref1}, the authors used wireless networks to implement scalable QML on quantum states. The key features of their research were the creation of the first QFL dataset and the development of the associated framework that combined \texttt{TensorFlow Quantum} and \texttt{TensorFlow Federated}. The performance of their framework was rigorously tested in the presence of noise and adversarial attacks, and in both cases, the framework was robust and indicated good performance.

In \cite{ref2}, the research proposed a framework for training hybrid quantum-classical classifiers in NISQ-era computers while adhering to the properties of FedML, namely privacy, and distributing the computational loads across devices. It was demonstrated that the proposed framework maintains high model accuracy while converging quickly when benchmarked against non-federated learning cases.

In \cite{ref3}, the research provided a comprehensive overview of QFL and described, in detail, the underlying concepts, principles, taxonomy of QFL techniques, and emerging applications in the form of a review-to-date paper. Comparably, the work in \cite{ref6} presented congruent discussions. 

In \cite{ref4}, the authors generically discussed the integration of QML into the realm of DNNs, emphasizing the potential of quantum computers to achieve exponential speedups over their classical counterparts for classes of complex problems. In particular, the research highlighted the apparent security advantages of QC, stemming from the principle of no-cloning, which prevents arbitrary unknown quantum states from being copied. By acknowledging the current limitations of NISQ-era quantum hardware, the authors proposed utilizing variational quantum algorithms (VQA) to facilitate computation with a limited number of qubits. Additionally, the paper delved into leveraging noise to tackle basic ML tasks advantageously. The contributions introduced include successfully implementing differentially-private federated training on hybrid quantum-classical models, achieving comprehensive security of the learning process by joining merits from two distinct privacy-preserving classical techniques, and presenting a viable privacy preservation method for QML on NISQ devices. This demonstrated the effectiveness of differentially-private QFL in mitigating privacy concerns while maintaining competitiveness. 

In \cite{ref7}, the authors explored an application area of QFL within the security domain for autonomous vehicles via quantum-inspired optimization. The research introduced several novelties: Adversarial defense in FedML and enhanced resilience against antipathetical attacks.  

In \cite{ref8}, the research introduced a QFL framework, \texttt{QuantumFed}, which demonstrated that multiple quantum nodes with local quantum data could collaborate and train a global model together without directly pooling the data into a central server. The framework was shown to be robust, and overcame data privacy challenges by using QNNs. 

Several researchers have presented their unique algorithms and perspectives on QFL \cite{ref99, ref88, ref77}, exploring aspects like data governance, privacy, and dynamic adaptability in quantum federated learning. However, these studies often encounter challenges in diverse data handling and model flexibility under varying conditions. In contrast, our FedQNN framework addresses these limitations by effectively managing various datasets, including genomics and healthcare, and demonstrating consistently high accuracy across these datasets.

\section{FedQNN Framework\label{sec3}}   
This section discusses our proposed FedQNN framework that trains QNNs with federated learning settings, as described in Algorithm \ref{alg:federated_qnn}.
\begin{algorithm}[htpb]
\caption{Our Proposed FedQNN Framework}
\label{alg:federated_qnn}
\begin{algorithmic}[1]
\State \textbf{Input:}
\begin{itemize}
    \item $\text{max\_iterations}$: Maximum number of training iterations
    \item $\text{num\_clients}$: Number of client devices
    \item $\text{num\_data\_points}$: Total number of data points
    \item $\text{num\_features\_to\_use}$: Number of features to use
    \item $X$: Dataset features
    \item $y$: Dataset labels
    \item $\text{learning\_rate}$: Learning rate for the QNN optimization
    \item $\text{optimizer}$: Optimizer used for the QNN optimization
\end{itemize}
\State \textbf{Initialization:}
\State initialize hyperparameters and dataset
\State Split the dataset into client-specific data
\State Initialize the QNN parameters on each client device

\For{$\text{iteration} \gets 1$ to $\text{max\_iterations}$}
  \State \textbf{Federated Learning Round} $[\text{iteration}]$
  \For{$i \gets 1$ to $\text{num\_clients}$}
    \State \textbf{Client Training Phase (Client $i$)}
    \State Train QNN on local client data using $\text{learning\_rate}$ and $\text{optimizer}$
    \State Communicate the updated QNN parameters from client $i$ to the server
  \EndFor

  \State \textbf{Global Model Update Phase}
  \State Aggregate the QNN parameters from all client devices on the server
  \State Update the global QNN parameters using the aggregated QNN parameters from all client devices

  \State \textbf{Evaluation Phase}
  \State Calculate the accuracy and loss on the test dataset using the global QNN.
  \State Store the accuracy and loss for the current iteration
\EndFor
\State \textbf{Output:} Accuracy, loss, and other metrics.
\end{algorithmic}
\end{algorithm}
\subsection{QNN Architecture}
This proposed QNN model, which we have developed, is inspired by foundational works within the domain \cite{refc,refqnn}. It begins by encoding classical data into quantum states through angle embedding. This process involves rotations about the $X$-axis, transforming classical data features into quantum equivalents. The QNN circuit is then established, consisting of a series of quantum gates such as rotations about the $Y$-axis, Controlled-NOT (CNOT), and Hadamard ($H$) gates. These gates are crucial for state manipulation, entanglement generation, and superposition creation, enabling the QNN to traverse an expansive computational landscape.
 \begin{figure}[htpb]
     \centering
     \includegraphics[width=1\linewidth]{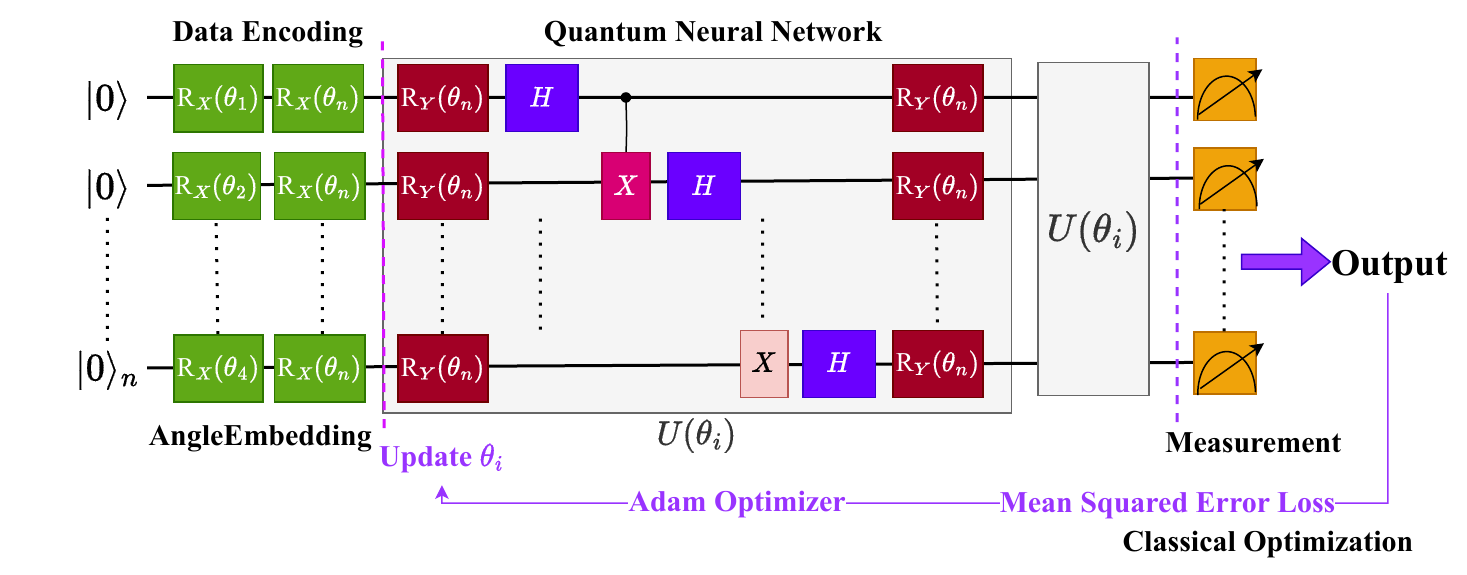}
     \caption{Architecture of the QNN used in this work. The process begins with data encoding through angle embedding, applying a series of $R_X(\theta_n)$ gates. These states are then processed by the QNN, which consists of alternating layers of $H$, $CNOT$, and parameterized rotation gates $R_Y(\theta_n)$. The unitary operations $U(\theta_i)$ further manipulate the quantum states. The output of the QNN is measured, and the result is used to calculate the MSE loss for model optimization. Parameter updates are performed iteratively using the Adam optimizer, adjusting the rotation angles to minimize the loss function, thereby refining the model with each iteration.}
     \label{qnn}
 \end{figure}
Upon constructing the QNN circuit as presented in Fig. \ref{qnn}, measurement operations yield classical outcomes from quantum states; this represents the QNN's output. Mathematically, this is represented as
\begin{equation}
\ket{\psi} = U(\boldsymbol{\theta})\ket{0}, \quad y_{\text{predicted}} = \langle\psi|M|\psi\rangle,
\end{equation}
where $\ket{\psi}$ is the quantum state post-embedding, $U(\boldsymbol{\theta})$ is the unitary operation representing the QNN circuit parameterized by $\boldsymbol{\theta}$, and $M$ is the measurement operator.

\subsection{QNN Training and Optimization}
The QNN undergoes training; this is an iterative process where the parameters $\boldsymbol{\theta}$ are adjusted to minimize the Mean Squared Error (MSE) cost function, often utilizing classical optimization techniques like the Adam optimizer. This optimization enhances the QNN's ability to predict or represent data accurately.

In the network, the quantum gates serve the following purpose: The $H$ gate creates superposition by mapping $\ket{0}\to\left(\ket{0}+\ket{1}\right)/\sqrt{2}$, the rotation gates also create superposition by rotating about the $X$- and $Y$-axes respectively, and introduce phase shifts. The $U(\boldsymbol{\theta})$ block serves as the optimization phase where the network parameters are iteratively adjusted to obtain optimal values. The measurement apparatus, technically not gates, ascertain the state of the output from the optimization block. This choice of architecture efficiently processes the data and repetitively adjusts the network parameters such that the loss is minimized. 

The MSE cost function is expressed as
\begin{equation}
\text{MSE} = \frac{1}{N}\sum_{i=1}^{N}(y_i - y_{\text{predicted},i})^2,
\end{equation}
where $y_i$ represents the true labels, and $y_{\text{predicted},i}$ are the predictions from the QNN.

\subsection{FedQNN Framework Design}
In the proposed FedQNN framework, each client device independently maintains its local data, as illustrated in Fig. \ref{qfl}. Instead of centralizing the data, the quantum model updates -- encapsulated as quantum parameters -- are communicated to a central server. The server then amalgamates these updates, forming a global quantum model to enhance overall model performance and accuracy.
The aggregation of parameters from different clients is mathematically described as
\begin{equation}
\boldsymbol{\theta}_{\text{global}}\overset{.}{=}\boldsymbol{\Theta} = \frac{1}{K}\sum_{k=1}^{K}\boldsymbol{\theta}_k,
\end{equation}
where $\boldsymbol{\theta}_k$ are the parameters from the $k^{\text{th}}$ client, and $K$ is the total number of clients.
\begin{figure*}[htbp]
    \centering   
    \includegraphics[width=1\linewidth]{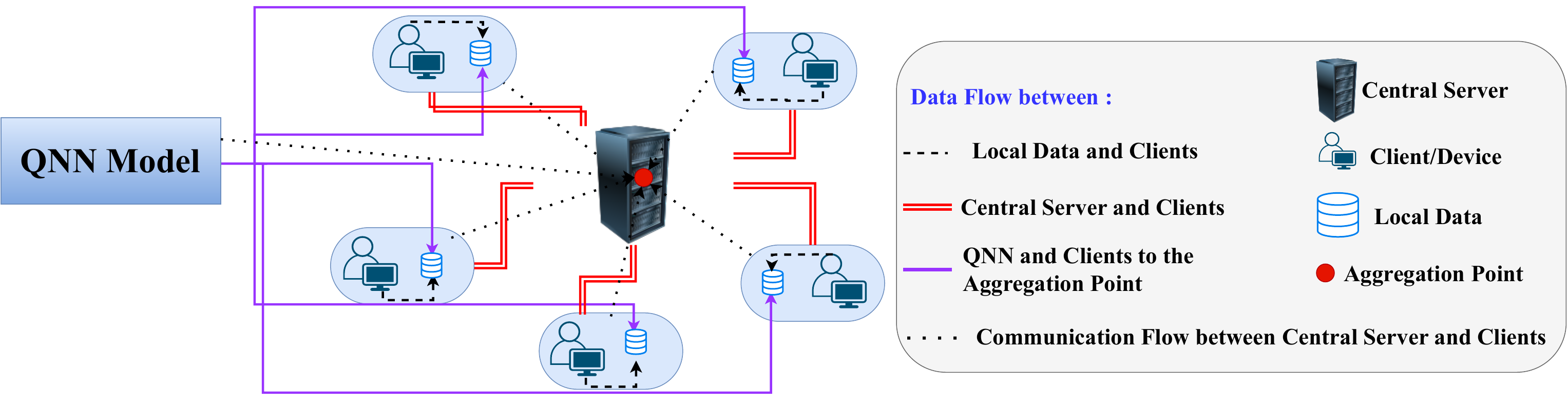}
    \caption{The FedQNN framework where each \textbf{Client/Device} retains its \textbf{Local Data}, ensuring data privacy as no raw data is shared with the \textbf{Central Server}. Clients independently train a \textbf{QNN model} with their data, and only quantum model updates--comprising parameters or operations--are communicated to the central server, which functions as an \textbf{Aggregation Point}, synthesizing a global model from these updates. Secure communication protocols are employed for exchanging updates and preventing the exposure of individual data. }
    \label{qfl}
\end{figure*}

\subsection{Secure Communication and Collaborative Learning}
The proposed FedQNN framework incorporates secure communication protocols such as privacy-preserving aggregation algorithms and uses quantum gates for encryption, ensuring that clients can share quantum updates securely while protecting individual client data. This method enables the network to benefit from all clients' collective intelligence and learning without compromising data privacy.
Additionally, our FedQNN method, integrating QNNs with federated learning principles, represents a significant advancement in combining QC's computational strengths with the practicalities of handling sensitive and distributed data in ML.

\section{Results and Discussion\label{sec4}}
\subsection{Experimental Setup}
In our research, as shown in Fig. \ref{tools} and in Tab. \ref{hyperparameters}, we implement our QFL algorithm using \texttt{PennyLane} -- A Python library with \texttt{Pytorch} backend, known for its QML capabilities \cite{pennylane}. Our computational setup, provided by Google Colab, includes a system equipped with dual CPUs and an NVIDIA Tesla T4 GPU. We conduct extensive experimentation across three diverse datasets: The Iris dataset \cite{irisdataset}, the breast cancer dataset \cite{breastcancer}, and a synthetic DNA dataset designed for classifying promoter and non-promoter sequences. Each dataset presents unique challenges and characteristics. The Iris dataset, with its 150 samples of three different Iris flower species, tests basic classification abilities. The breast cancer dataset consists of 286 instances with nine attributes, including linear and nominal types for both classes. The synthetic DNA dataset, tailored for genomics applications, stands out for its focus on the binary classification of genetic sequences. It contains 200 samples, categorized into promoters and non-promoters.
\begin{figure}
    \centering
    \hspace{-0.5cm}
    \includegraphics[width=1\linewidth]{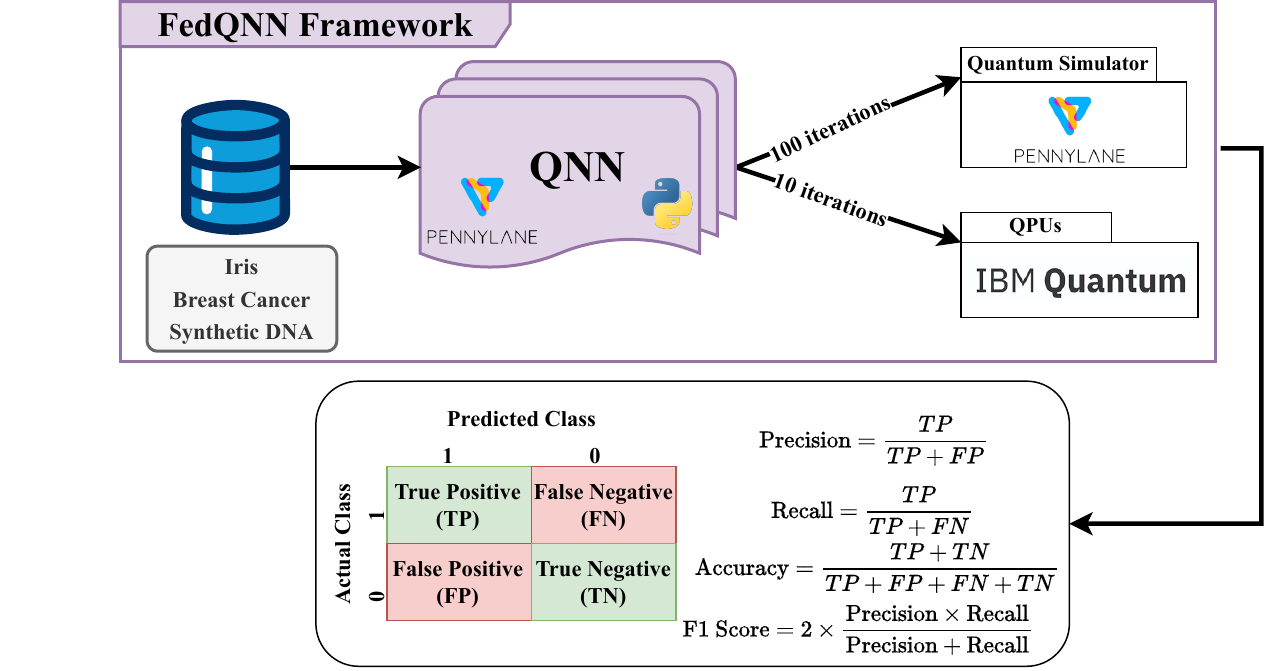}
    \caption{Experimental setup and tool flow for conducting the experiments: The process begins with datasets fed into the QNN within the FedQNN framework. The QNN, implemented using the PennyLane framework, undergoes optimization through iterative training, with the number of iterations indicated for the Pennylane simulator and IBM QPUs. Post-training, the model's performance is assessed using standard binary classification metrics, including precision, recall, accuracy, and the F1 score.}
    \label{tools}
\end{figure}
\begin{table}[htpb]
\centering
\caption{Hyperparameters and configurations used in our FedQNN framework.}
\label{hyperparameters}
\begin{tabular}{ll}
\toprule
\textbf{Parameter}& \textbf{Value}\\
\midrule
Number of Qubits& 4\\
Optimizer& Adam\\
Step Size for Adam Optimizer& 0.1\\
Max Iterations for Local Training& 100\\
Parameters for QNN& Randomly initialized, Size: 16\\
Number of Clients & 1, 2, 3, 4, 5\\
Max Iterations for Global Training& 100\\
Test Set Size& 20\% of the dataset\\
Accuracy Metric& Binary classification accuracy\\
Loss Metric& MSE\\
\bottomrule
\end{tabular}
\end{table}
Throughout experimentation, we observed intriguing accuracy dynamics over training iterations throughout the analysis with FedQNN, across the three diverse datasets. The accuracy plots shown in Fig. \ref{accuracies} for each dataset show a distinct zigzag pattern, reflecting the convergence and fluctuations in the learning process. However, it is important to highlight that the mean accuracy over ten distinct trials, with each trial comprising 100 iterations, did not exhibit this zigzag behavior for the three datasets, suggesting a more stable learning trend upon averaging multiple iterations. Despite the variations, each model consistently reaches an impressive peak accuracy, typically between 85\% and 90\%. This remarkable level of accuracy serves as compelling evidence of the effectiveness of our FedQNN framework in multi-dataset classification scenarios. Tab. \ref{tab} supports these findings, showcasing essential metrics such as precision, recall, F1-score, and accuracy for each dataset. The consistently high accuracy values reiterate the robustness and reliability of our method for handling various datasets.
\begin{figure*}[htbp]
    \centering
    \begin{subfigure}{.30\linewidth}
        \centering
                \begin{tikzpicture}
            \node[anchor=south west, inner sep=0] (image) at (0,0) {\includegraphics[width=\linewidth]{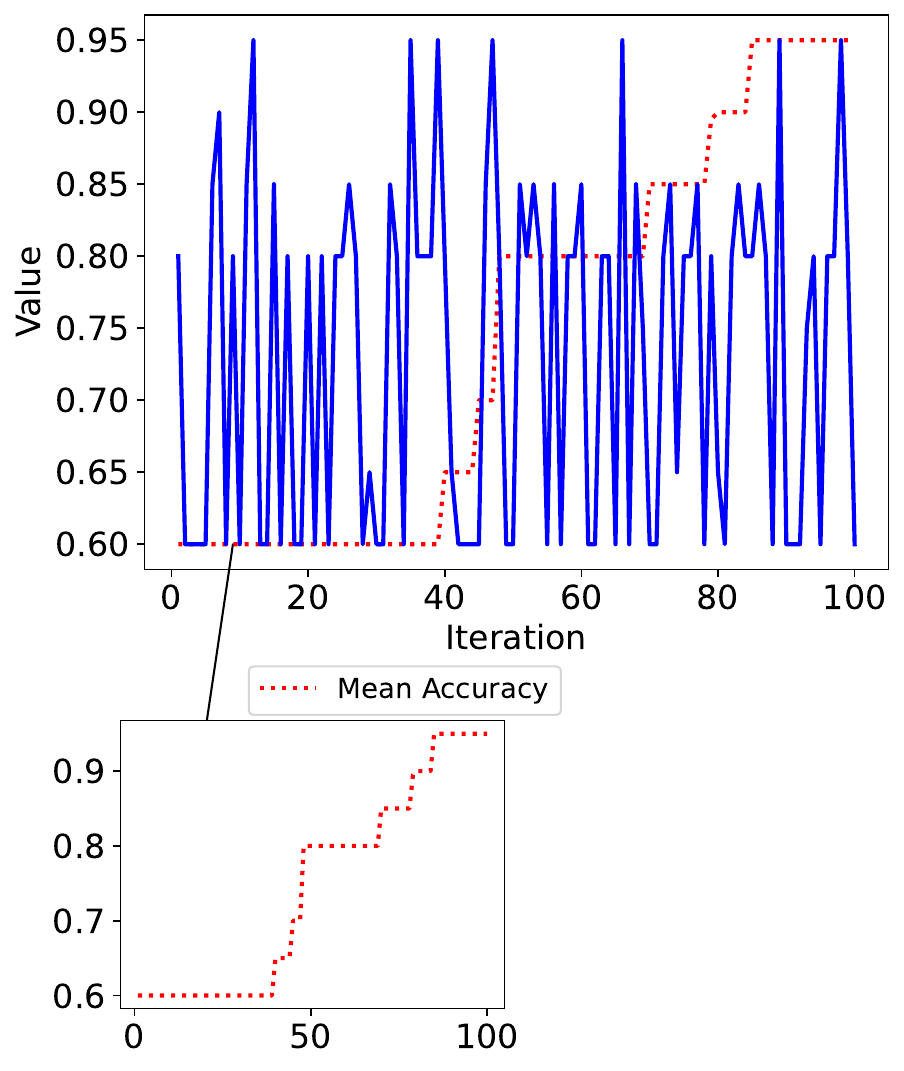}};
                \node[anchor=south west, font=\bfseries\Large, text=black] at (4.5,0.4) {(a)};

            \begin{scope}[x={(image.south east)},y={(image.north west)}]
                \node[circle, draw=purple, fill=white, inner sep=1pt] (pointerB) at (0.207,1-0.09) {\textcolor{purple}{1}};
                \draw[-latex, purple, thick] (pointerB) -- +(0.07, 0.052);
                \node[circle, draw=blue, fill=white, inner sep=1pt] (pointerB) at (0.51,1-0.76) {\textcolor{blue}{4}};
                \draw[-latex, blue, thick] (pointerB) -- +(0.0, 0.078);                
            \end{scope}
        \end{tikzpicture}
        \label{iris1}
    \end{subfigure}%
    \begin{subfigure}{.291\linewidth}
        \centering
             \begin{tikzpicture}
            \node[anchor=south west, inner sep=0] (image) at (0,0) {\includegraphics[width=\linewidth]{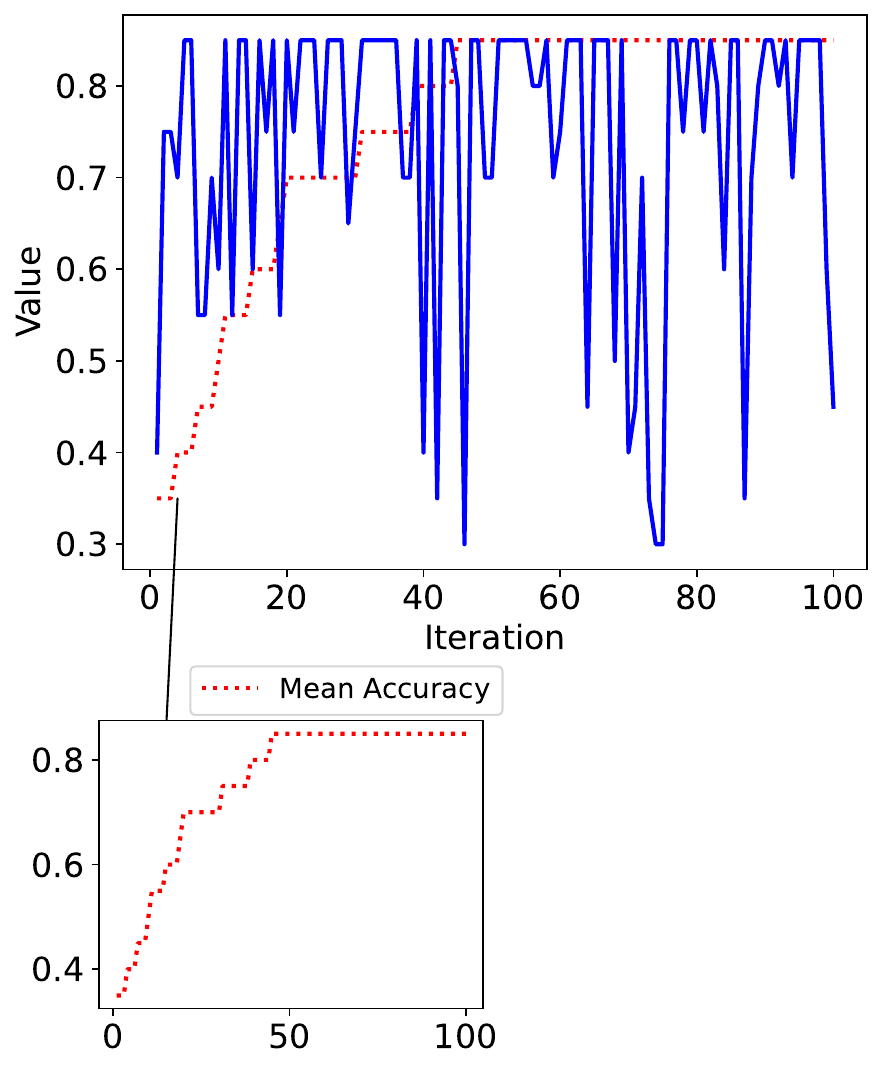}};
    \node[anchor=south west, font=\bfseries\Large, text=black] at (4.5,0.4) {(b)};
            \begin{scope}[x={(image.south east)},y={(image.north west)}]
                \node[circle, draw=purple, fill=white, inner sep=1pt] (pointerB) at (0.59,1-0.2) {\textcolor{purple}{2}};
                \draw[-latex, purple, thick] (pointerB) -- +(-0.01, 0.16);
                \node[circle, draw=blue, fill=white, inner sep=1pt] (pointerB) at (0.51,1-0.76) {\textcolor{blue}{5}};
                \draw[-latex, blue, thick] (pointerB) -- +(0.0, 0.078); 
            \end{scope}
        \end{tikzpicture}        
        \label{cancer1}
    \end{subfigure}
    \begin{subfigure}{.291\linewidth}
        \centering
             \begin{tikzpicture}
            \node[anchor=south west, inner sep=0] (image) at (0,0) {\includegraphics[width=\linewidth]{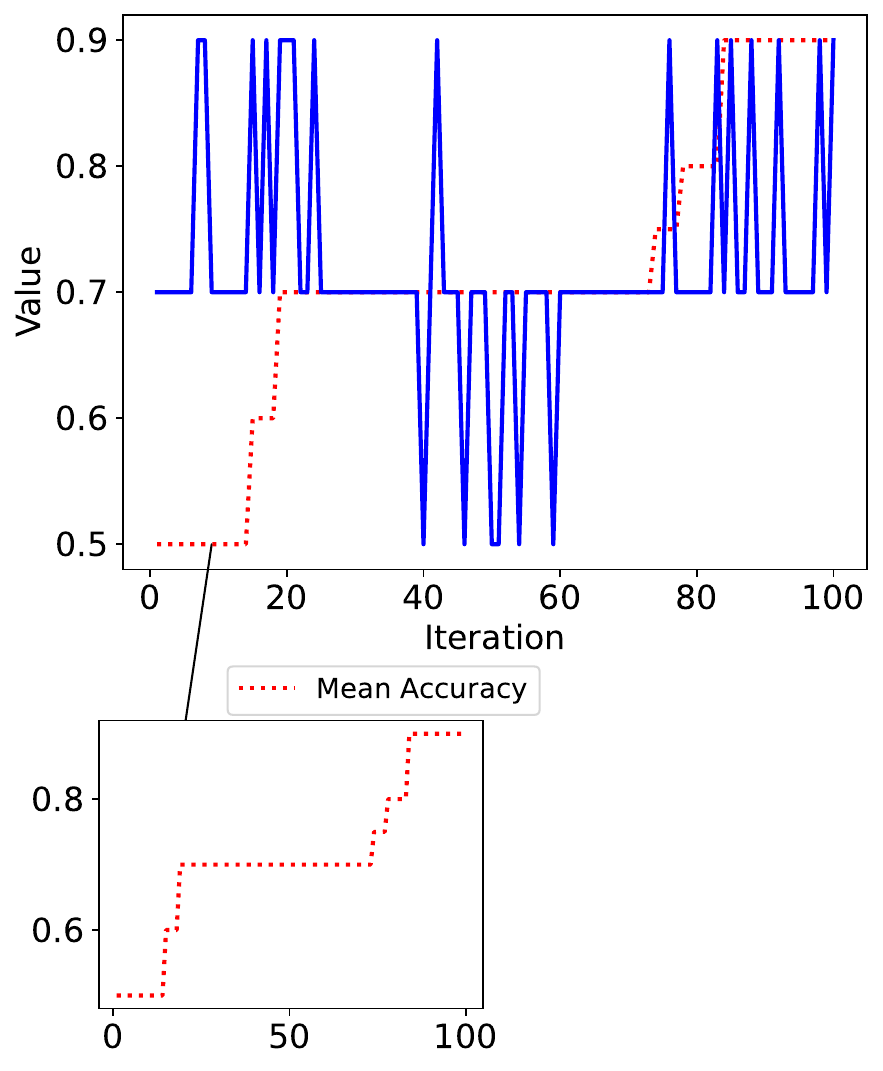}};
        \node[anchor=south west, font=\bfseries\Large, text=black] at (4.5,0.4) {(c)};
            \begin{scope}[x={(image.south east)},y={(image.north west)}]
                \node[circle, draw=purple, fill=white, inner sep=1pt] (pointerB) at (0.43,1-0.09) {\textcolor{purple}{3}};
                \draw[-latex, purple, thick] (pointerB) -- +(0.063, 0.052);
                \node[circle, draw=blue, fill=white, inner sep=1pt] (pointerB) at (0.51,1-0.76) {\textcolor{blue}{6}};
                \draw[-latex, blue, thick] (pointerB) -- +(0.0, 0.078); 
            \end{scope}
        \end{tikzpicture}          
        \label{dna1}
    \end{subfigure}
    \caption{Accuracies over iterations for various datasets: \textbf{(a)} Iris, \textbf{(b)} breast cancer, and \textbf{(c)} DNA, displaying a distinctive zigzag pattern where high accuracies are periodically reached, as exemplified by pointers \textcircled{\raisebox{-0.9pt}{1}}, \textcircled{\raisebox{-0.9pt}{2}}, and \textcircled{\raisebox{-0.9pt}{3}}. This pattern, consistent across the three datasets, suggests a recurrent peaking in model performance at certain iterations. When considering the mean accuracy--calculated from ten trials--a clearer learning curve emerges. This curve demonstrates a general trend of improvement and eventual stabilization in model performance, highlighted by pointers \textcircled{\raisebox{-0.9pt}{4}}, \textcircled{\raisebox{-0.9pt}{5}}, and \textcircled{\raisebox{-0.9pt}{6}} towards the end of the iterations.}
    \label{accuracies}
\end{figure*}
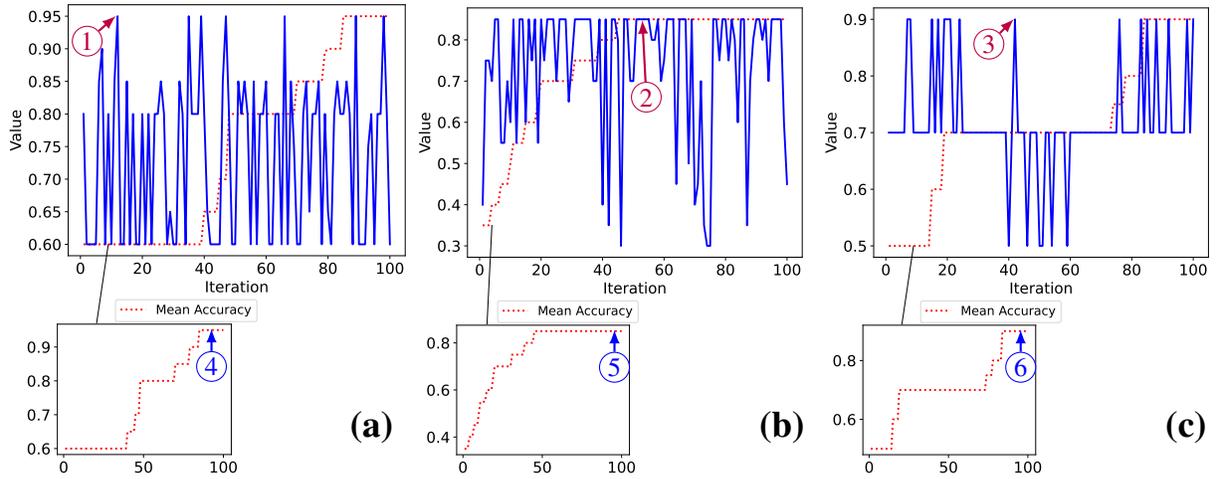

\begin{table}[htbp]
    \centering
    \caption{Classification reports for different datasets.}
    \begin{tabular}{lcccc}
        \toprule
        \multirow{2}{*}{Dataset} & \multicolumn{4}{c}{Metrics} \\ 
        \cmidrule(lr){2-5}
        & Precision & Recall & F1-Score & Accuracy\\
        \midrule
        Iris & 0.89 & 0.89 & 0.89 & 0.9 \\
        \midrule
        Breast Cancer & 0.85 & 0.85 & 0.92 &  0.86  \\
        \midrule
        DNA & 0.9 &0.88 & 0.89 &  0.9 \\
        \bottomrule
    \end{tabular}
    \label{tab}
\end{table}
The proposed FedQNN experiments explore the intriguing relationship between the number of clients and model accuracy. The results in Fig. \ref{clients} demonstrate that increasing the number of clients can lead to remarkable improvements in accuracy. This phenomenon highlights the power of collaborative learning in a federated setting, where each client contributes to the collective intelligence, while keeping its data private. The robustness of FedQNN across different datasets reaffirms its potential as an effective framework for classification tasks.
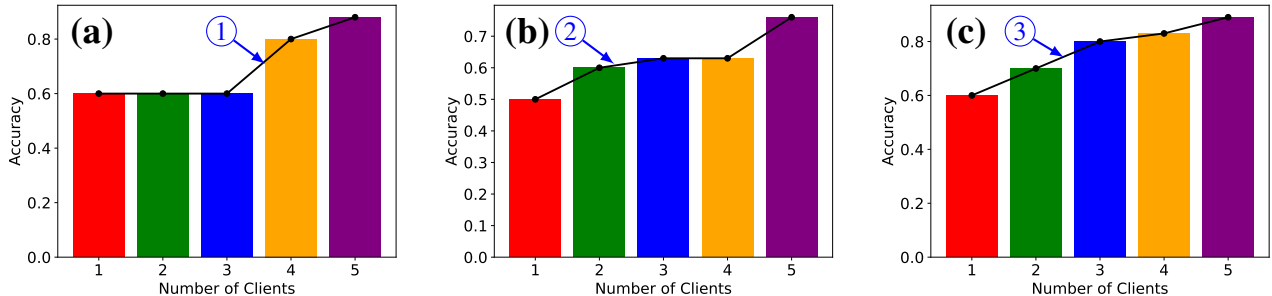
\begin{figure*}[htbp]
    \centering
    \begin{subfigure}{.32\linewidth}
        \centering
        \begin{tikzpicture}
            \node[anchor=south west, inner sep=0] (image) at (0,0) {\includegraphics[width=\linewidth]{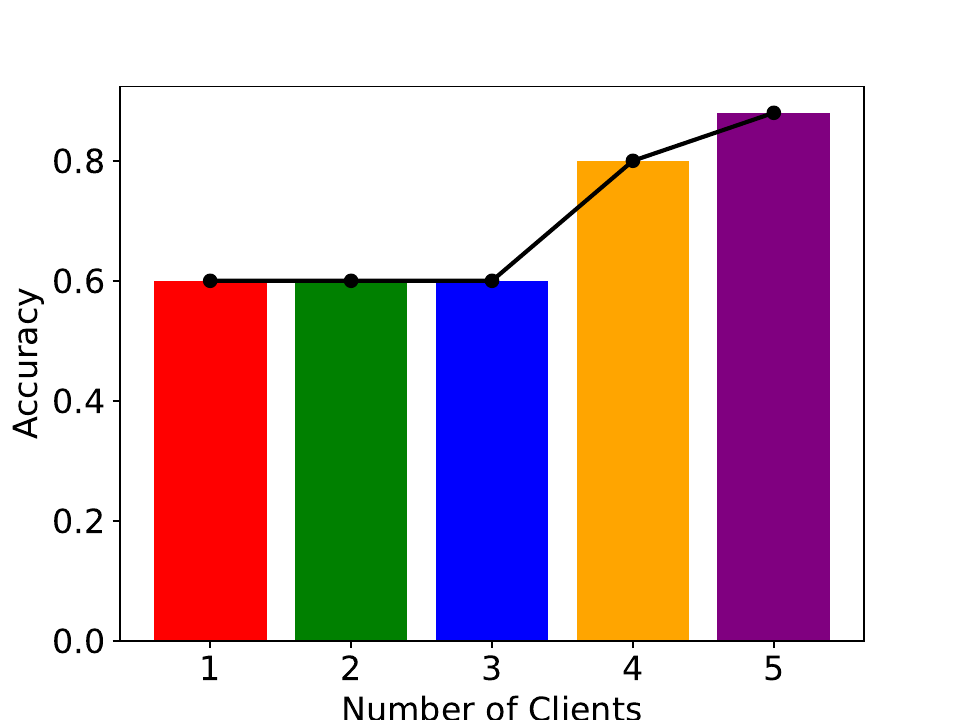}};
               \node[anchor=north east, font=\bfseries\Large, text=black] at (1.6,3.8) {(a)};            
            \begin{scope}[x={(image.south east)},y={(image.north west)}]
                \node[circle, draw=blue, fill=white, inner sep=1pt] (pointerA) at (0.5,0.8) {\textcolor{blue}{1}};
                \draw[-latex, blue, thick] (pointerA) -- +(0.1, -0.1);
            \end{scope}
        \end{tikzpicture}
        \label{iris2}
    \end{subfigure}%
    \begin{subfigure}{.32\linewidth}
        \centering
        \begin{tikzpicture}
            \node[anchor=south west, inner sep=0] (image) at (0,0) {\includegraphics[width=\linewidth]{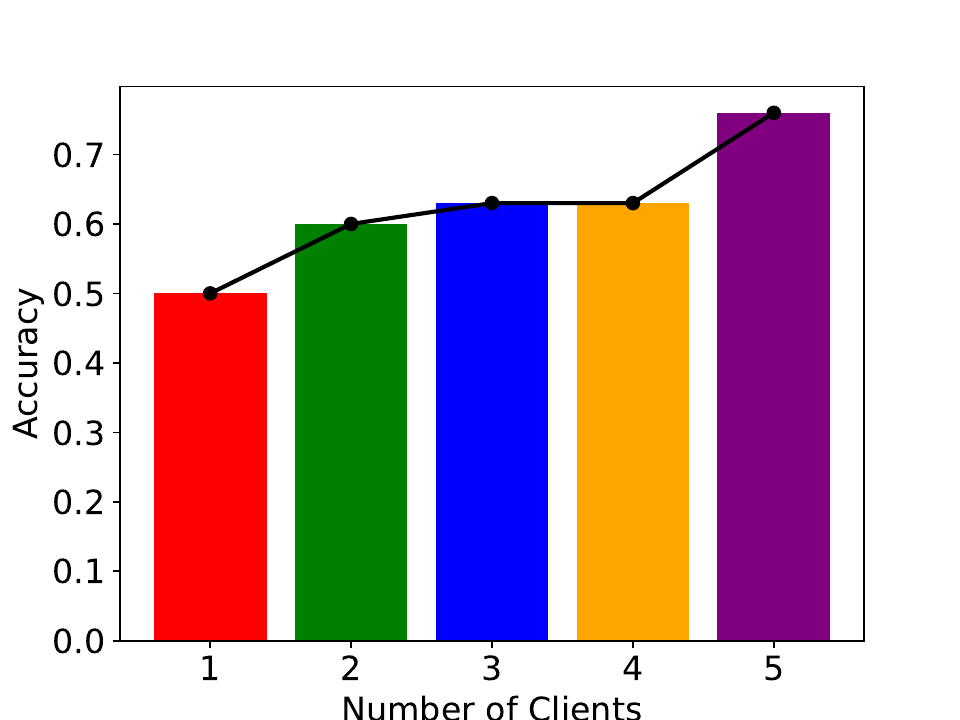}};
               \node[anchor=north east, font=\bfseries\Large, text=black] at (1.6,3.8) {(b)};               
            \begin{scope}[x={(image.south east)},y={(image.north west)}]
                \node[circle, draw=blue, fill=white, inner sep=1pt] (pointerB) at (0.3,0.8) {\textcolor{blue}{2}};
                \draw[-latex, blue, thick] (pointerB) -- +(0.1, -0.1);
            \end{scope}
        \end{tikzpicture}
        \label{can2}
    \end{subfigure}%
    \begin{subfigure}{.32\linewidth}
        \centering
        \begin{tikzpicture}
            \node[anchor=south west, inner sep=0] (image) at (0,0) {\includegraphics[width=\linewidth]{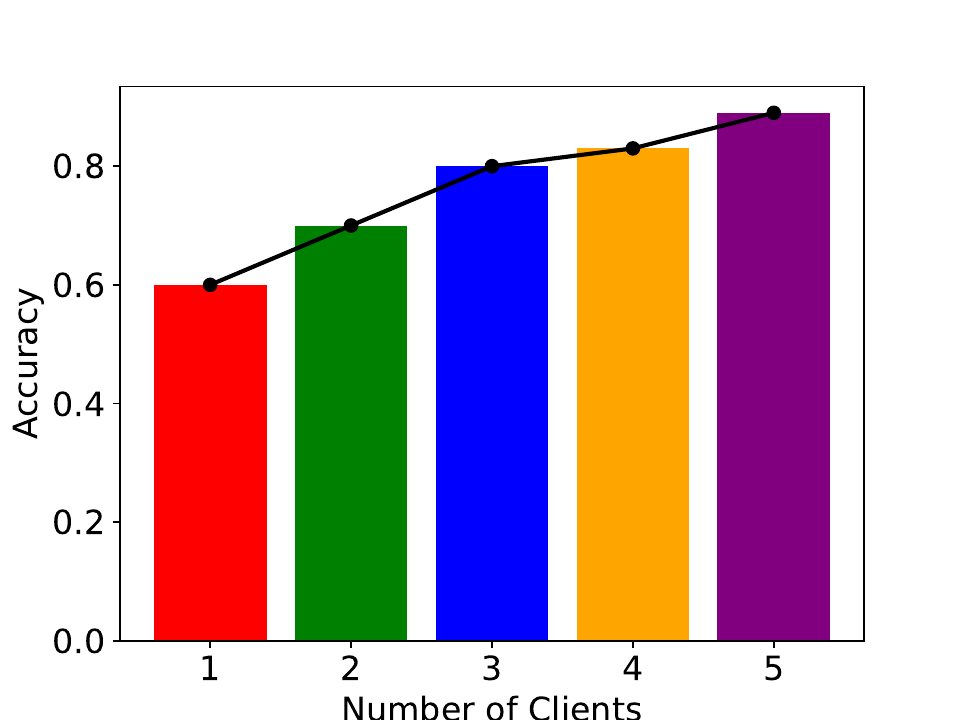}};
               \node[anchor=north east, font=\bfseries\Large, text=black] at (1.6,3.8) {(c)};               
            \begin{scope}[x={(image.south east)},y={(image.north west)}]
                \node[circle, draw=blue, fill=white, inner sep=1pt] (pointerC) at (0.33,0.8) {\textcolor{blue}{3}};
                \draw[-latex, blue, thick] (pointerC) -- +(0.1, -0.08);
            \end{scope}
        \end{tikzpicture}
        \label{dna2}
    \end{subfigure}
    \caption{Accuracies over the number of clients for various datasets:  \textbf{(a)} showcases the Iris dataset, where the accuracy starts at 0.6 with one client and shows a marked increase to 0.88 with five clients, as indicated by the trajectory \textcircled{\raisebox{-0.9pt}{1}}, \textbf{(b)} presents the breast cancer dataset, and it shows a gradual improvement in accuracy from 0.5 with a single client to 0.76 with five clients, though the rate of increase plateaus between three and four clients, as highlighted by the trajectory \textcircled{\raisebox{-0.9pt}{2}}, and \textbf{(c)} presents the DNA dataset with accuracy rising from 0.6 to 0.89 as the number of clients increases from one to five, with the trajectory \textcircled{\raisebox{-0.9pt}{3}} emphasizing the consistent trend. These results underscore the positive correlation between the number of contributing clients and accuracy.}
    \label{clients}
\end{figure*}

Furthermore, in Fig. \ref{qpu}, we present comprehensive tests across three different QPUs from IBM Quantum: \texttt{ibm\_nairobi}, \texttt{ibm\_lagos}, and \texttt{ibm\_perth} \cite{ibm}. Notably, all three QPUs deliver efficient results, with accuracy rates surpassing the 80\% threshold. However, it is worth highlighting that these accuracy scores did exhibit some degree of fluctuation, which can be attributed to factors such as quantum noise and decoherence, variations in gate fidelity, discrepancies in quantum volume, and sensitivity to environmental conditions. Each of these factors can contribute to subtle yet impactful differences in performance across the QPUs. Importantly, these evaluations are conducted with only $10$ iterations to test the model's rapid convergence, further indicating the potential of our model to excel in QC settings, even with limited refinement, and the promising prospects of its adaptability and performance on various QPUs.
\begin{figure}
    \centering

        \begin{tikzpicture}
            \node[anchor=south west, inner sep=0] (image) at (0,0) {\includegraphics[width=1\linewidth]{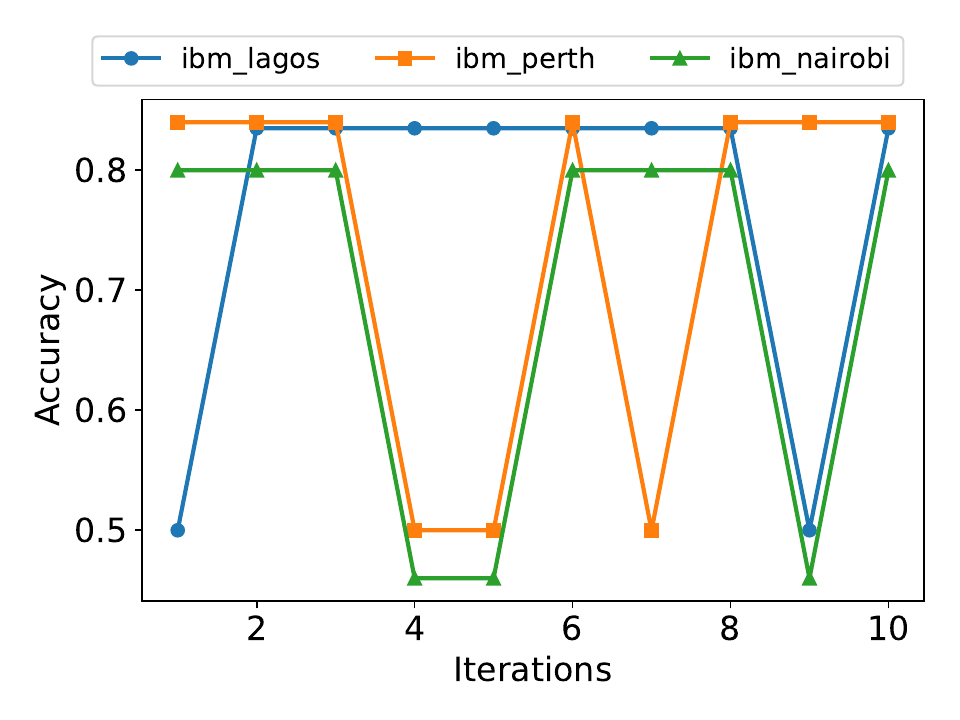}};
            \begin{scope}[x={(image.south east)},y={(image.north west)}]
                \node[circle, draw=blue, fill=white, inner sep=1pt] (pointerB) at (0.5,0.7) {\textcolor{blue}{1}};
                \draw[-latex, blue, thick] (pointerB) -- +(0, 0.125);
                \node[circle, draw=blue, fill=white, inner sep=1pt] (pointerB) at (0.755,0.22) {\textcolor{blue}{2}};
                \draw[-latex, blue, thick] (pointerB) -- +(0.09, 0.04);
               \node[circle, draw=blue, fill=white, inner sep=1pt] (pointerB) at (0.35,0.29) {\textcolor{blue}{3}};
                \draw[-latex, blue, thick] (pointerB) -- +(0.075, -0.02);     
               \node[circle, draw=blue, fill=white, inner sep=1pt] (pointerB) at (0.35,0.22) {\textcolor{blue}{4}};
                \draw[-latex, blue, thick] (pointerB) -- +(0.08, -0.02);                
            \end{scope}
        \end{tikzpicture}
    \caption{The accuracy of the model using QPUs from IBM Quantum, for ibm\_lagos, a peak accuracy of 0.835 is consistently observed from \textcircled{\raisebox{-0.9pt}{1}} onward, with a notable dip back to 0.50 at \textcircled{\raisebox{-0.9pt}{2}}. For ibm\_perth, accuracy starts with a value of 0.84, experiences a drop to 0.5 at \textcircled{\raisebox{-0.9pt}{3}}, and fluctuates between these values, finishing strong with an accuracy of 0.84, and for ibm\_nairobi, a similarly high accuracy of 0.8, drops to 0.46 at \textcircled{\raisebox{-0.9pt}{4}}, and then mirrors the pattern of IBM perth ending at 0.8.}
    \label{qpu}
\end{figure}
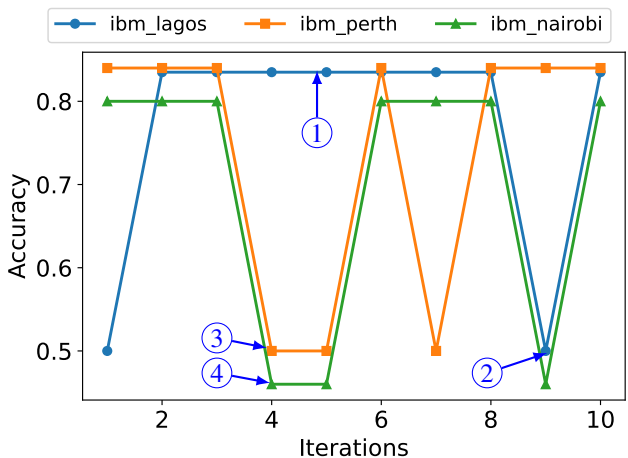
\subsection{Discussion}
The comprehensive experimentation and analysis of the proposed FedQNN framework yield significant insights, underlining its potential in QML. A critical observation is the framework's adeptness in managing a spectrum of datasets, illustrating its adaptability and robustness. This versatility is pivotal in QML, where applications often span diverse data types and problem domains.

Notably, the framework exhibits stable learning dynamics over multiple iterations despite initial accuracy fluctuations. Such a trend indicates a resilient learning algorithm capable of consistent performance enhancement over time. This aspect is crucial in practical scenarios, where models must adapt and stabilize despite variable data inputs.

Performance metrics across the datasets consistently register high, highlighting the model's precision, recall, and accuracy proficiency. These metrics are crucial benchmarks in ML, and their high values in this framework reinforce its efficacy in balanced classification tasks.

The study also reveals the positive influence of increasing client collaboration within the federated network. This finding accentuates the benefits of a federated approach in ML, where diversity in data sources enriches the learning process and enhances model accuracy.

The experiments, across different QPUs from IBM Quantum, demonstrate the framework's adaptability and efficient performance, even with limited iterations. This adaptability to various quantum hardware architectures is a testament to the framework's potential for widespread application in QC settings.

In a comparative analysis of QFL methods as presented in Tab. \ref{comparison}, our FedQNN framework demonstrates outstanding performance across various datasets; when evaluated on the CIFAR-10 dataset for planes versus cars, a hybrid quantum-classical classifier (HQCC) achieved an accuracy of 94.05\% with just two local epochs of federated training \cite{ref2}. Additionally, the \texttt{SlimQFL} and \texttt{Vanilla QFL} models achieve accuracies of 77\% and 76\%, respectively, on the mini-MNIST dataset 
\cite{ref88}. Even when dealing with non-Independent and Identically Distributed (IID) data on MNIST-3, a QNN model reached a 70\% accuracy mark \cite{ref77}. Notably, our results show superior performance on three diverse datasets: Iris (90\%), breast cancer (86\%), and DNA (90\%), which is competitive with \texttt{qFedInf} and \texttt{qFedAvg} that report accuracies up to 92.7\% and 88.4\% on MNIST-2, and 75.4\% and 66.7\% on Fashion-MNIST \cite{refc2}. These results underscore the fruitful utility of our QFL framework and its versatility in handling different types of data, thus marking a significant step forward in QFL research.

The overall outcomes from this investigation signify a promising avenue for QML, particularly in its application across varied datasets and quantum environments. However, these successes also pave the way for future explorations, particularly optimizing quantum circuit designs and enhancing integration techniques within federated learning systems. Moreover, the results suggest potential broader implications in fields requiring rigorous data privacy and security, positioning QFL as a valuable tool in privacy-sensitive collaborative research and applications.
\begin{table}[htpb]
\centering
\caption{Comparison of QFL frameworks on different datasets.}
\label{comparison}
\begin{tabular}{llll}
\toprule
Reference & Dataset & Key Features & Accuracy \\
\midrule
\cite{ref2} & CIFAR-10 & HQCC & 94.05\% \\
\cite{ref88} & Mini-MNIST  & SlimQFL & 77\% \\
& & Vanilla QFL & 76\% \\
\cite{ref77} & MNIST-3  & QNN & 70\% \\
\cite{refc2} & MNIST-2 & qFedInf & \(92.7 \pm 0.2\)\% \\
& & qFedAvg & \(88.4 \pm 0.8\)\% \\
& Fashion-MNIST & qFedInf & \(75.4 \pm 0.3\)\% \\
& & qFedAvg & \(66.7 \pm 1.3\)\% \\
\textbf{Our FedQNN} & Iris &  Distributed QNN & \textbf{90\%} \\
& Breast Cancer &  & \textbf{86\%} \\
& DNA & & \textbf{90\%} \\
\bottomrule
\end{tabular}
\end{table}

\section{Conclusion\label{sec5}}
Our investigation into QFL as a secure and accurate data classification method across decentralized and encrypted datasets has revealed immense potential. By deploying a QNN with a FedQNN framework, we have demonstrated the ability to highlight the benefits of group learning while safeguarding individual data privacy.

Experiments spanning diverse datasets, from genomics to medicine, confirmed the adaptability and effectiveness of our FedQNN framework. The accuracy consistently exceeds 85\%, proving our framework's reliability in classifying data from multiple sources. Furthermore, the proposed approach exhibits impressive scalability, maintaining high accuracy even with increasing client participation, highlighting its suitability for large-scale collaborative learning tasks. This resilience extends to challenging conditions as we successfully test our FedQNN framework on real QPUs from IBM, achieving over 80\% accuracy, and thereby solidifying its real-world applicability.

In conclusion, FedQNN emerges as a powerful tool for secure and accurate collaborative data classification, opening new doors for privacy-focused research and development across various fields. By advancing communication efficiency and exploring more complex quantum algorithms, we can unlock the full potential of QFL as a transformative technology for secure and collaborative data-driven intelligence. We note that there are several challenges with QFL in general, such as: 
\begin{enumerate}
\item The fragility of quantum states, thereby making them error-prone, resulting in the loss of information.
\item The effect of noise and the introduction of instabilities.
\item Using classical aggregation methods, like averaging, is not the best method for quantum state agglomeration.
\item Client-side models on NISQ-era hardware introduce erraticity into the models.
\end{enumerate}
However, this research lays the foundation for a future where data privacy and collective knowledge can drive breakthroughs in diverse areas, from healthcare to material science, ushering in a new era of secure and collaborative innovation.

\section*{Acknowledgment}
This work was supported in part by the NYUAD Center for Quantum and Topological Systems (CQTS), funded by Tamkeen under the NYUAD Research Institute grant CG008.


\end{NoHyper}

\begin{thebibliography}{00}
\bibitem{refa} Zaman, K., Marchisio, A., Hanif, M. A., \& Shafique, M. (2023). \textit{A Survey on Quantum Machine Learning: Current Trends, Challenges, Opportunities, and the Road Ahead}. arXiv preprint arXiv:2310.10315.

\bibitem{refb} Bharti, K., Cervera-Lierta, A., Kyaw, T. H., Haug, T., Alperin-Lea, S., Anand, A., ... \& Aspuru-Guzik, A. (2022). \textit{Noisy intermediate-scale quantum algorithms}. Reviews of Modern Physics, \textbf{94}(1), 015004.

\bibitem{refc} Beer, K., Bondarenko, D., Farrelly, T., Osborne, T. J., Salzmann, R., Scheiermann, D., \& Wolf, R. (2020). \textit{Training deep quantum neural networks}. Nature communications, \textbf{11}(1), 808.

\bibitem{refd} Kashif, M., \& Al-Kuwari, S. (2024). \textit{ResQNets: a residual approach for mitigating barren plateaus in quantum neural networks}. EPJ Quantum Technology, \textbf{11}(1), 4.

\bibitem{refd1} Zaman, K., Ahmed, T., Hanif, M. A., Marchisio, A., \& Shafique, M. (2024). \textit{A Comparative Analysis of Hybrid-Quantum Classical Neural Networks}. arXiv preprint arXiv: 2402.10540.  

\bibitem{refd2} Zaman, Kamila, et al. (2024). \textit{Studying the Impact of Quantum-Specific Hyperparameters on Hybrid Quantum-Classical Neural Networks}. arXiv preprint arXiv:2402.10605. 

\bibitem{refd3} Maouaki, Walid El, et al. (2024). \textit{AdvQuNN: A Methodology for Analyzing the Adversarial Robustness of Quanvolutional Neural Networks}. arXiv preprint arXiv:2403.05596. 

\bibitem{refd4} Kashif, M., \& Shafique, M. (2024). \textit{ResQuNNs: Towards Enabling Deep Learning in Quantum Convolution Neural Networks}. arXiv preprint arXiv:2402.09146. 

\bibitem{refd5} Kashif, Muhammad, et al. (2024). \textit{Alleviating barren plateaus in parameterized quantum machine learning circuits: Investigating advanced parameter initialization strategies}. arXiv preprint arXiv: 2311.13218. 

\bibitem{ref5} MacMahan, H. B., Moore, E., Ramage, D., Hampson, S., \& y Arcas, B. A. (2016). \textit{Communication-Efficient Learning of Deep Networks from Decentralized Data}. arXiv preprint arXiv:1602.05629.

\bibitem{refe} Li, W., Lu, S., \& Deng, D. L. (2021). \textit{Quantum federated learning through blind quantum computing}. Science China Physics, Mechanics \& Astronomy, \textbf{64}(10), 100312.

\bibitem{reff} Xia, Q., \& Li, Q. (2021). \textit{Quantumfed: A federated learning framework for collaborative quantum training}. In 2021 IEEE Global Communications Conference (GLOBECOM) (pp. 1-6). IEEE.

\bibitem{refc1} Chen, L., Xue, K., Li, J., Li, R., Yu, N., Sun, Q., \& Lu, J. (2023). \textit{Q-DDCA: Decentralized Dynamic Congestion Avoid Routing in Large-Scale Quantum Networks}. IEEE/ACM Transactions on Networking.

\bibitem{refy} Chehimi, M., Chen, S. Y. C., Saad, W., Towsley, D., \& Debbah, M. (2023). \textit{Foundations of quantum federated learning over classical and quantum networks}. IEEE Network.

\bibitem{ref2} Chen, S. Y-C., \& Yoo, S. (2021). \textit{Federated Quantum Machine Learning}. Entropy, \textbf{23}(4), pp. 1-14.

\bibitem{ref9} Innan, N., Khan, M. A. Z., \& Bennai, M. (2023). \textit{Financial Fraud Detection: A Comparative Study of Quantum Machine Learning Models}. International Journal of Quantum Information.
\bibitem{ref10} Innan, N. et al. (2023). \textit{Financial Fraud Detection Using Quantum Graph Neural Networks}. Quantum Machine Intelligence, \textbf{6}(1), 1-18.
\bibitem{ref11} Innan, N., Khan, M. A. Z., \& Bennai, M. (2023). \textit{Quantum Computing for Electronic Structure Analysis: Ground State Energy and Molecular Properties Calculations}. Materials Today Communications, \textbf{38}.
\bibitem{reftt} Ullah, U., \& Garcia-Zapirain, B. (2024). \textit{Quantum Machine Learning Revolution in Healthcare: A Systematic Review of Emerging Perspectives and Applications}. IEEE Access.

\bibitem{reftt2} Khan, M. A. Z., Innan, N., Galib, A. A. O., \& Bennai, M. (2024). \textit{Brain Tumor Diagnosis Using Quantum Convolutional Neural Networks}. arXiv preprint arXiv:2401.15804.

\bibitem{ref12} Innan, N., Khan, \& M. A. Z. (2023). \textit{Classical-to-Quantum Sequence Encoding in Genomics}. arXiv preprint arXiv: 2304.10786.

\bibitem{ref14} Innan, N. et al. (2023). \textit{Quantum State Tomography Using Quantum Machine Learning}. arXiv preprint arXiv:2308.10327.

\bibitem{ref100} Feynman, R. (1982). \textit{Simulating Physics with Computers}. International Journal of Theoretical Physics, \textbf{21}(6/7), pp. 467-488.

\bibitem{qubit} Dejpasand, M. T., \& Sasani Ghamsari, M. (2023). \textit{Research trends in quantum computers by focusing on qubits as their building blocks}. Quantum Reports, \textbf{5}(3), 597-608.

\bibitem{ref1} Chehimi, M., \& Saad, W. (2022). \textit{Quantum Federated Learning with Quantum Data}. IEEE International Conference on Acoustics, Speech and Signal Processing (ICASSP), Singapore,  pp. 8617-8621.

\bibitem{ref3} Ren, C., Yu, H., Yan, R., Xu, M., Shen, Y., Zhu, H., Niyato, D., Dong, Z. Y., Kwek, L. C. (2023). \textit{Towards Quantum Federated Learning}. arXiv preprint arXiv:2306.09912.

\bibitem{ref6} Larasati, H. T., Firdaus, M., \& Kim, H. (2022). \textit{Quantum Federated Learning: Remarks and Challenges}. IEEE 9th International Conference on Cyber Security and Cloud Computing (CSCloud)/2022 IEEE 8th International Conference on Edge Computing and Scalable Cloud (EdgeCom), China, pp. 1-5.

\bibitem{ref4} Rofougaran, R., Yoo, S., Tseng, H-H., \& Chen, S, Y-C. (2023). \textit{Federated Quantum Machine Learning with Differential Privacy}. arXiv preprint arXiv:2310.06973.

\bibitem{ref7} Yamany, W., Moustafa, N., \& Turnbull, B. (2021). \textit{OQFL: An Optimized Quantum-Based Federated Learning Framework for Defending Against Adversarial Attacks in Intelligent Transportation Systems}. IEEE Transactions on Intelligent Transportation Systems, \textbf{24}(1), pp. 893-903.

\bibitem{ref8} Xia, Q., \& Li, Q. (2021). \textit{QuantumFed: A Federated Learning Framework for Collaborative Quantum Training}. IEEE Global Communications Conference (GLOBECOM), Spain, pp. 1-6. 

\bibitem{ref99} Bhatia, A. S., Kais, S., \& Alam, M. A. (2023). \textit{Federated quanvolutional neural network: a new paradigm for collaborative quantum learning}. Quantum Science and Technology, \textbf{8}(4), 045032.

\bibitem{ref88} Yun, W. J., Kim, J. P., Jung, S., Park, J., Bennis, M., \& Kim, J. (2022). \textit{Slimmable quantum federated learning}. arXiv preprint arXiv:2207.10221.

\bibitem{ref77} Zhang, Y., Zhang, C., Zhang, C., Fan, L., Zeng, B., \& Yang, Q. (2022). \textit{Federated Learning with Quantum Secure Aggregation}. arXiv preprint arXiv:2207.07444.

\bibitem{refqnn} Schuld, M., Sinayskiy, I., \& Petruccione, F. (2014). \textit{The quest for a quantum neural network}. Quantum Information Processing, \textbf{13}, 2567-2586.

\bibitem{pennylane} Bergholm, V., Izaac, J., Schuld, M., Gogolin, C., Ahmed, S., Ajith, V.,... \& Killoran, N. (2018). \textit{Pennylane: Automatic differentiation of hybrid quantum-classical computations}. arXiv preprint arXiv:1811.04968.

\bibitem{irisdataset} Fisher, R. A. (1988). \textit{Iris}. UCI Machine Learning Repository, \url{https://doi.org/10.24432/C56C76}.

\bibitem{breastcancer} Zwitter, Matjaz, and Soklic, Milan. (1988). \textit{Breast Cancer}. UCI Machine Learning Repository, \url{https://doi.org/10.24432/C51P4M}.

\bibitem{ibm} IBM Quantum, \url{https://quantum.ibm.com/}.

\bibitem{refc2} Zhao, H. (2023). \textit{Non-IID quantum federated learning with one-shot communication complexity}. Quantum Machine Intelligence, \textbf{5}(1), 3.


\end{thebibliography}
\end{document}